\documentclass[11pt]{amsart}
\usepackage{amsmath}
\usepackage{amssymb}
\usepackage{amsthm} 
\usepackage{bbm}
\usepackage{graphicx}
\usepackage{algorithm}
\usepackage{algorithmic}
\usepackage{clrscode}
\usepackage{fullpage}

\newtheorem{cor}{Corollary}
\newtheorem{thm}{Theorem}
\newtheorem{lem}{Lemma}

\newcommand{\polylog}{\mathop{\rm polylog}}
\DeclareMathOperator*{\argmin}{arg\,min}

\begin{document}
\pagestyle{plain}
\newenvironment{frcseries}{\fontfamily{frc} \selectfont}{}
\newcommand{\textfrc}[1]{{\frcseries #1}}
\newcommand{\mathfrc}[1]{\text{\textfrc{#1}}}

\title{Compressed Sensing with Sparse Binary Matrices:  Instance Optimal Error Guarantees in Near-Optimal Time}
%
\author{M. A. Iwen \\
Duke University, Box 90320 \\
Durham, NC 27708-0320\\
Email:  markiwen@math.duke.edu}

\maketitle
\thispagestyle{empty}

\begin{abstract}
A compressed sensing method consists of a rectangular measurement matrix, $M \in \mathbbm{R}^{m \times N}$ with $m \ll N$, together with an associated recovery algorithm, $\mathcal{A}:  \mathbbm{R}^m \rightarrow \mathbbm{R}^N$.  Compressed sensing methods aim to construct a high quality approximation to any given input vector ${\bf x} \in \mathbbm{R}^N$ using only $M {\bf x} \in \mathbbm{R}^m$ as input.  In particular, we focus herein on instance optimal nonlinear approximation error bounds for $M$ and $\mathcal{A}$ of the form $\left \| {\bf x} - \mathcal{A} \left( M {\bf x} \right) \right \|_p \leq \left\| {\bf x} - {\bf x}^{\rm opt}_k \right\|_p + C k^{1/p - 1/q} \left\| {\bf x} - {\bf x}^{\rm opt}_k \right\|_q$ for ${\bf x} \in \mathbbm{R}^N$, where ${\bf x}^{\rm opt}_k$ is the best possible $k$-term approximation to ${\bf x}$.

In this paper we develop a compressed sensing method whose associated recovery algorithm, $\mathcal{A}$, runs in $O\left( (k \log k ) \log N \right)$-time, matching a lower bound up to a $O(\log k)$ factor.  This runtime is obtained by using a new class of sparse binary compressed sensing matrices of near optimal size in combination with sublinear-time recovery techniques motivated by sketching algorithms for high-volume data streams.  The new class of matrices is constructed by randomly subsampling rows from well-chosen incoherent matrix constructions which already have a sub-linear number of rows.  As a consequence, fewer random bits  than previously required are needed in order to select the rows utilized by the fast reconstruction algorithms considered herein.
\end{abstract}

\section{Introduction}
\label{sec:intro}

Noisy group testing problems generally involve designing pooling schemes which use as few expensive tests as possible in order to identify a small number of important elements from a large universe, $\mathcal{U}$, of items (see, e.g., \cite{du2000combinatorial}).  In this setting each one of the expensive tests in question corresponds to observing the result of an experiment, or calculation, performed on a different subset of $\mathcal{U}$.  If each test is sufficiently sensitive to the small number of hidden items in $\mathcal{U}$ that must be identified, one might hope that testing a correspondingly small number of subsets of $\mathcal{U}$ in bulk would still allow the hidden elements to be found.  Thus, designing a pooling scheme corresponds to choosing a good collection of subsets of $\mathcal{U}$ to observe so that tests performed on these subsets will always allow one to discover a small number of important elements hidden within $\mathcal{U}$.  

Many data mining tasks can be cast in a similar framework -- that is -- as problems concerned with identifying a small number of interesting items from a tremendously large group without exceeding certain resource constraints (e.g., without using too much memory, communication power, runtime, etc.).  Specific examples include the sketching and monitoring of heavy-hitters in high-volume data streams \cite{charikar2002finding,cormode2003s}, source localization in sensor networks \cite{zheng2006nonadaptive}, and the design of high throughput sequencing schemes for biological specimen analysis \cite{erlich2009dna}.  Note that pooling schemes in many such group testing related applications naturally correspond to binary matrices (i.e., because each row of the binary matrix, ${\bf r} \in \{ 0,1 \}^N$, selects a subset for testing/observation).  Furthermore, it is generally better for these binary binary matrices to have a small number of nonzero entries in each column (i.e., because this reduces the number of times each item in $\mathcal{U}$ must be tested/observed).  Thus, we focus on designing sparse binary measurement matrices herein.

Roughly speaking, one can cast many such applications as a type of compressed sensing \cite{donohoCS} problem.  The large set containing the small number of important elements we want to identify is modeled as a vector, ${\bf x} \in \mathbbm{R}^N$.  The $n^{th}$ entry in the vector, $x_n$, is a real number which indicates the ``importance'' of the $n^{th}$ set element (the larger the magnitude, the more important).  Our goal is now to locate $k \ll N$ of the largest magnitude entries of ${\bf x}$ (i.e., the important elements).  Unfortunately, for reasons that vary with the specific problem at hand (e.g., because only $o(N)$ memory is available in the massive data stream context), we are allowed to store just $m \ll N$ linear measurements of ${\bf x}$ which we must compute during a single pass over its entries.  The $m$ linear measurement operators are represented as a measurement matrix, $M \in \mathbbm{R}^{m \times N}$.  Having access to only $M {\bf x} \in \mathbbm{R}^m$, we seek to identify, and then estimate, the $k$ largest magnitude entries of ${\bf x}$.  This identification and estimation is performed by a sparse recovery algorithm, $\mathcal{A}:  \mathbbm{R}^m \rightarrow \mathbbm{R}^N$, which (implicitly) returns a vector in $\mathbbm{R}^N$ having $O(k)$ nonzero entries.  We prefer $\mathcal{A}$ to be fast, especially for applications involving massive data sets.

In this paper we consider the design of sparse matrices $M \in \{ 0, 1 \}^{m \times N}$, with $m \ll N$, together with associated nonlinear functions, $\mathcal{A}:  \mathbbm{R}^m \rightarrow \mathbbm{R}^N$, which have the property that $\mathcal{A} \left( M{\bf x} \right) \approx {\bf x}$ for all vectors ${\bf x} \in \mathbbm{R}^N$  that are well approximated by their best $k$-term approximation,
\begin{equation}
{\bf x}^{\rm opt}_k := \argmin_{{\bf y} \in \mathbbm{R}^N, \| {\bf y} \|_0 \leq k} \| {\bf x} - {\bf y} \|_2.\footnote{Here $\| {\bf y} \|_0$ denotes the number of nonzero entries in ${\bf y } \in \mathbbm{R}^N$, while $\| {\bf y} \|_p$ denotes the standard $\ell_p$-norm for all $p \geq 1$, i.e., $\| {\bf y} \|_p = \left(\sum^{N-1}_{n=0} |y_n|^p \right)^{1/p}$ for all ${\bf y} \in \mathbbm{R}^N$.}
\end{equation}
More specifically, we will focus on designing $(M,\mathcal{A})$-pairs which achieve error guarantees of the form 
\begin{equation}
\left \| {\bf x} - \mathcal{A} \left( M {\bf x} \right) \right \|_p \leq  \min_{{\bf y} \in \mathbbm{R}^N, \| {\bf y} \|_0 \leq k} \left\| {\bf x} - {\bf y} \right\|_p + C_{p,q} \cdot k^{1/p - 1/q} \left\| {\bf x} - {\bf y} \right\|_q
\label{equ:lplq}
\end{equation}
for constants $1 \leq q \leq p \leq 2$, and $C_{p,q} \in \mathbbm{R}^+$ (e.g., see \cite{BestkTerm, HolgerCSBook}).  We will refer to such an error guarantee as an ``$\ell_p, \ell_q$'' error guarantee below.  

Over the past several years this type of design problem has achieved a considerable amount of attention under the moniker of ``compressed sensing'' (e.g., see \cite{HolgerCSBook,donohoCS,CRTCS,devore2007deterministic,HardThreshforCS,berinde2008combining,jafarpour2009efficient}, and references therein).  Most compressed sensing papers -- this one included -- generate their measurement matrices, $M$, randomly.  This leads to two different probabilistic models in which the aforementioned ``$\ell_p,\ell_q$'' error guarantees may hold.   In the first model, a single randomly generated measurement matrix, $M$, is shown to satisfy \eqref{equ:lplq} \textit{for all} ${\bf x} \in \mathbbm{R}^N$ with high probability.  We will refer to this as the ``for all'' model.  In the second model, a randomly generated measurement matrix is shown to satisfy \eqref{equ:lplq} \textit{for each given} ${\bf x} \in \mathbbm{R}^N$ with high probability (assuming that $M$ is generated independently of ${\bf x}$).  We will refer to this second model as the ``for each'' model.  All results proven herein are proven in the second, ``for each'', model.

\subsection{Results and Related Work}
\label{sec:IntroResandRel}

Any sparse recovery algorithm, $\mathcal{A}$, that achieves either an ``$\ell_1,\ell_1$'', ``$\ell_2,\ell_1$'', or ``$\ell_2,\ell_2$'' error guarantee in the ``for each'' model must use an associated measurement matrix, $M$,  having at least $m \geq C k \log (N / k) )$ rows \cite{do2010lower,price20111+}.\footnote{$C$ will always represent an absolute constant.}  Note that this implies an $\Omega(k \log (N / k))$ lower runtime complexity bound for the recovery algorithm, $\mathcal{A}$.  It remains an open problem to prove (or disprove) the existence of a $O(k \log N)$-time recovery algorithm achieving any such error guarantee.  
In this paper we present a compressed sensing matrix/recovery algorithm pair, $(M,\mathcal{A})$, with an ``$\ell_2, \ell_1$'' guarantee, where $\mathcal{A}$ runs in $O((k \log k) \log N)$-time $-$ a single $O(\log k)$-factor from the known lower bound.  We also present two other compressed sensing results which can be obtained using the same methods:  one which uses an optimal number (up to constant factors) of randomly selected rows from an incoherent binary matrix as measurements, and another $O(k \log^2 N)$-time recovery result which requires fewer random bits\footnote{More precisely, the number of random bits is $O(\log^2 k)$.  To the best of our knowledge this represents the first fast recovery result which requires a number of random bits that is entirely independent of $N$, the length of ${\bf x}$.} than previous algorithms (i.e., less randomness). 

Previous work involving the development of compressed sensing methods having both sub-linear time reconstruction algorithms, and the type of ``$\ell_p,\ell_q$'' error guarantees considered herein, began with \cite{cormode2006combinatorial}.  In \cite{cormode2006combinatorial} Cormode et al. built on streaming algorithm techniques with weaker error guarantees (e.g., see \cite{charikar2002finding,cormode2003s,cormode2005improved}) in order to develop $O(k \log^3 N)$-time recovery algorithms, $\mathcal{A}$, with associated ``$\ell_2,\ell_2$'' error guarantees in the ``for each" model.  Similar techniques were later utilized by Gilbert et. al. in \cite{gilbert2010approximate} to create sub-linear time algorithms with the same error guarantees, but whose associated measurement matrices, $M \in \mathbbm{R}^{m \times N}$, have a near-optimal number of rows up to constant factors (i.e., $m = O(k \log N)$).  Other related compressed sensing methods with fast runtimes and ``$\ell_2,\ell_1$'' error guarantees in the ``for all" model were also considered in \cite{Gilbert:2007}.  Unlike these previous methods, the compressed sensing methods developed herein utilize the combinatorial properties of a new class of sparse binary measurement matrices formed by randomly selecting sub-matrices from larger incoherent matrices.

Perhaps the measurement matrices considered herein are most similar to previous compressed sensing matrices based on unbalanced expander graphs (see, e.g., \cite{berinde2008combining,indyk2008near,jafarpour2009efficient}).  Indeed, the measurement matrices used in this paper are created by randomly sampling rows from larger binary matrices that are, in fact, the adjacency matrices of a subclass of unbalanced expander graphs.  However, unlike previous approaches which use the properties of general unbalanced expanders, we use different combinatorial techniques which allow us to develop $O(k \polylog N)$-time recovery algorithms.  To the best of our knowledge, the runtimes we obtain by doing so are the best known for any such method having ``$\ell_p,\ell_q$'' error guarantees.

See Figure~\ref{tab:Compare1} for a comparison of the sub-linear time compressed sensing results proven herein (last two rows) with previous sub-linear time compressed sensing results discussed above (first three rows).  The columns of Figure~\ref{tab:Compare1} list the following characteristics of each compressed sensing method:  $(i)$ the number of measurement matrix rows, $m$, $(ii)$ the runtime complexity of the recovery algorithm, and $(iii)$ the ``$\ell_p,\ell_q$'' error guarantee achieved by the method.  All error guarantees hold in the ``for each'' model unless indicated otherwise by a $\checkmark$.  

\begin{figure}
\begin{center}
\begin{tabular}{|c|c|c|c|}
\hline
\textbf{Paper} & \textbf{Measurements}, $m$ & \textbf{Runtime of $\mathcal{A}$} & \textbf{Error Guarantee}\\
\hline
\cite{Gilbert:2007} & $k \log^{\geq 2} N$ & $k^2 \log^{\geq 2} N$ & $\ell_2,\ell_1$ \checkmark \\
\cite{cormode2006combinatorial} & $k \log^3 N$ & $k \log^3 N$ & $\ell_2, \ell_2$ \\
\cite{gilbert2010approximate} & $k \log N$ & $k \log^{\geq 2} N$ & $\ell_2,\ell_2$\\
\hline
Herein$^*$ & $k \log^2 N$ & $k \log^2 N$ & $\ell_2, \ell_1$ \\
Herein & $(k \log k) \log N$ & $(k \log k) \log N$ & $\ell_2,\ell_1$\\
\hline
\end{tabular}
\end{center}
\tiny{\checkmark Error guarantees hold in the ``for all'' model.}\\
\tiny{$^*$Requires only $O(\log^2 k)$ random bits.}
\caption{Summary of previous sub-linear time results, and the results obtained herein.}
\label{tab:Compare1}
\end{figure}

\subsection{Techniques and Organization}

It has been shown that all binary matrices satisfying easily verifiable coherence conditions\footnote{Any matrix whose maximum inner product between all pairs of columns is small compared to the minimal number of ones in each column satisfies the required coherence conditions.  See Section~\ref{sec:prelim} for details.} have strong combinatorial properties capable of producing entirely deterministic compressed sensing algorithms requiring $\Omega(k^2 \log N )$ runtime and measurements \cite{SIAM_Fourier_Matrix}.  In this paper we demonstrate a general means for utilizing these same types of matrices to construct compressed sensing approximation schemes with near-optimal runtime and sampling complexities.  Our new compressed sensing matrices are formed by randomly sampling a small number of rows from any sufficiently incoherent binary matrix.  The resulting random sub-matrices are then shown to still satisfy sufficiently strong combinatorial properties with respect to any given input vector, ${\bf x}$, in order to allow standard fast compressed sensing techniques (i.e., similar of those utilized in \cite{cormode2006combinatorial}) to produce accurate results.  Furthermore, the theory is developed in a modular fashion, making it easy to utilize different binary incoherent matrix constructions in order to generate new results.  We take advantage of this modularity in order to generate the two new results listed in Figure~\ref{tab:Compare1}, as well as to show that our new measurement matrices also allow for compressive sensing with an optimal number of measurements (up to constant factors) in $O(N \log N)$-time.\footnote{See Theorem~\ref{appthm:NewRecoverAlg} for details.} 
Each result is produced by utilizing a different combination of two incoherent binary matrix constructions:  deterministic algebraic constructions due to DeVore \cite{devore2007deterministic}, and randomly constructed incoherent binary matrices with fewer rows constructed below in Section~\ref{sec:OptExistence}.

The remainder of this paper is organized as follows:  In Section~\ref{sec:prelim} we fix notation and review existing results that are needed for later sections.  In Section~\ref{sec:OptExistence} we construct incoherent binary matrices with a near optimal number of rows. These new binary matrices ultimately allow the development of our $O\left( (k \log k ) \log N \right)$-time recovery result via the techniques developed later in Section~\ref{sec:SampMat}.  Section~\ref{sec:SampMat} constructs our compressed sensing measurement matrices by randomly sampling rows from the previously discussed binary incoherent matrices (i.e., from both the matrices reviewed in Section~\ref{sec:prelim} as well as the matrices constructed in Section~\ref{sec:OptExistence}).  Our main results are then proven in Section~\ref{sec:MainRes}.  Finally, we conclude with a short discussion in Section~\ref{sec:Conc}.

\section{Preliminaries}
\label{sec:prelim}

Let $[N] = \{ 0, \dots, N-1 \}$ for any $N \in \mathbbm{N}$.  We consider the elements of any given ${\bf x} \in \mathbbm{R}^N$ to be ordered according to magnitude by the sequence 
$j_0, j_1, \dots, j_{N-1} $ so that $ \left|x_{j_0} \right| \geq \left| x_{j_1} \right| \geq \dots \geq \left| x_{j_{N-1}} \right|.$
We set $S^{\rm opt}_{k} = \left\{ j_0, j_1, \dots, j_{k-1} \right\} \subset [N]$ for a given ${\bf x}$, and let ${\bf x}_{S^{\rm opt}_{k}} = {\bf x}^{\rm opt}_{k} \in \mathbbm{R}^N$ denote the associated vector with exactly $k$ nonzero entries:  
$$\left( x^{\rm opt}_{k} \right)_{j_0} = x_{j_0}, \left( x^{\rm opt}_{k} \right)_{j_1} = x_{j_1}, \dots, \left( x^{\rm opt}_{k} \right)_{j_{k-1}} = x_{j_{k-1}}.$$  
All results below deal with randomly sampling rows from a rectangular binary matrix whose columns are all nearly pairwise orthogonal.

\newtheorem{Definition}{Definition}
\begin{Definition}
Let $K, \alpha \in [N]$.  An $m \times N$ matrix, $M \in \{ 0,1 \}^{m \times N}$, is called $\left( K, \alpha \right)$-coherent if both of the following properties hold:
\begin{enumerate}
\item Every column of $M$ contains at least $K$ ones. 
\item For all $j, l \in [N]$ with $j \neq l$, the associated columns, $M_{\cdot,j} ~\textit{and}~ M_{\cdot,l} \in \{0,1 \}^m$, have $\left\langle M_{\cdot,j}, M_{\cdot,l} \right \rangle \leq \alpha$.
\end{enumerate}
\label{def:Coherent}
\end{Definition}
These matrices are closely related to nonadaptive group testing matrices, unbalanced expander graphs, binary matrices with the restricted isometry property, and codebook design problems in signal processing.  Several (implicit) constructions of $\left( K, \alpha \right)$-coherent matrices exist (e.g., the number theoretic and algebraic constructions of \cite{cormode2006combinatorial} and \cite{devore2007deterministic}, respectively).  In addition, every $\left( K, \alpha \right)$-coherent matrix must have $\Omega \left( \min \left\{ (K^2 / \alpha^2) \log_{K / \alpha} N, N \right\} \right)$ rows.  See \cite{SIAM_Fourier_Matrix} for details.

Given any binary matrix $M \in \{ 0, 1\}^{m \times N}$ with at least $K \in [m]$ ones in column $n \in [N]$, let $M(K,n)$ denote a $K \times N$ submatrix of $M$ created by selecting the first $K$ rows of $M$ with nonzero entries in the $n^{\rm th}$ column.  The following useful fact concerning $\left( K, \alpha \right)$-coherent matrices is proven in \cite{SIAM_Fourier_Matrix}.

\begin{lem}
Suppose $M$ is a $\left(K,\alpha \right)$-coherent matrix.  Let $n \in [N]$, $k \in \left[\frac{K}{\alpha} \right]$, $\epsilon \in (0,1]$, $c \in [2,\infty) \cap \mathbbm{N}$, and ${\bf x} \in \mathbbm{R}^{N}$.  If $K > c \cdot (k \alpha/ \epsilon)$ then $\left(M(K,n) \cdot {\bf x}\right)_j$ will be contained in the interval 
$\left( x_n - \frac{\epsilon \left\| {\bf x} - {\bf x}^{\rm opt}_{(k/\epsilon)} \right\|_1}{k}, ~x_n + \frac{\epsilon \left\| {\bf x} - {\bf x}^{\rm opt}_{(k/\epsilon)} \right\|_1}{k} \right)$ 
for more than $\frac{c-2}{c} \cdot K$ values of $j \in [K]$.
\label{applem:Mest}
\end{lem}

In addition to $\left( K, \alpha \right)$-coherent matrices, we will also utilize a \textit{bit-test matrix}, $\mathcal{B}_N \in \{ 0, 1 \}^{\left( 1 + \lceil \log_2 N \rceil \right) \times N}$, whose $n^{\rm th}$-column is a one followed by $n \in [N]$ written in base $2$.  These bit-test matrices will allow us to quickly identify large elements of a vector ${\bf x}$ using techniques from \cite{cormode2006combinatorial}.  The \textit{row tensor product} of two matrices, $\mathcal{A} \in \mathbbm{R}^{m_1 \times N}$ and $\mathcal{B} \in \mathbbm{R}^{m_2 \times N}$, denoted $\mathcal{A} \circledast \mathcal{B}$, is defined to be the $\left( m_1 \cdot m_2 \right) \times N$ matrix with entries are given by
$$\left(\mathcal{A} \circledast \mathcal{B}\right)_{i,j} = \mathcal{A}_{i~\textrm{mod}~m_1,j} \cdot \mathcal{B}_{\frac{\left(i - (i~\textrm{mod}~m_1)\right)}{m_1},j}.$$
The following Theorem was proven in \cite{SIAM_Fourier_Matrix}.

\begin{thm}
Let $\epsilon \in (0,1]$, $k \in \left[K \cdot \frac{\epsilon}{4 \alpha} \right]$, and ${\bf x} \in \mathbbm{R}^{N}$.  Furthermore, suppose that $M$ is an $m \times N$ binary matrix with the property that 
\begin{equation}
\left(M(K,n) \cdot {\bf x}\right)_j \in \left( x_n - \frac{\epsilon \left\| {\bf x} - {\bf x}^{\rm opt}_{(k/\epsilon)} \right\|_1}{k}, ~x_n + \frac{\epsilon \left\| {\bf x} - {\bf x}^{\rm opt}_{(k/\epsilon)} \right\|_1}{k} \right)
\label{eqn:CombProp}
\end{equation}
for more than $K / 2$ values of $j \in [K]$ for all $n \in [N]$.  Then, there exists an algorithm that takes $M$ and $\left( M \circledast \mathcal{B}_N \right) {\bf x}$ as input, and outputs a vector ${\bf z} \in \mathbbm{R}^N$ satisfying
$$ \left\| {\bf x} - {\bf z} \right\|_2 ~\leq~ \left\| {\bf x} - {\bf x}^{\rm~ opt}_k \right\|_2 + \frac{22 \epsilon \left\| {\bf x} - {\bf x}^{\rm opt}_{(k/\epsilon)} \right\|_1}{\sqrt{k}}.$$
Furthermore, the algorithm can be implemented to run in $O\left( m \log N \right)$ time.
\label{appthm:RecoverAlg}
\end{thm}

Pseudocode for a faster randomized variant of the algorithm referred to by Theorem~\ref{appthm:RecoverAlg} can be found in the Appendix.  This randomized variant and its associated measurement matrices are the focus of this paper.  Briefly put, both algorithms operate in two phases.  During the first phase all heavy entires of the input vector, ${\bf x}$, are identified using standard bit testing techniques \cite{cormode2006combinatorial}.  These heavy vector elements are then estimated during the second round using an approach along the lines of from the computer science streaming literature \cite{cormode2005improved,cormode2006combinatorial}.  The approximations provided by the binary matrices in Lemma~\ref{applem:Mest} guarantee that taking the median of all $K$ entires of $\left(M(K,n) \cdot {\bf x}\right)$ will provide a good estimate of each important entry, ${\bf x}_n$.


\subsection{Efficiently Storing a $\left( K, \left\lfloor \frac{\ln N}{\ln K} \right\rfloor \right)$-coherent Matrix}
\label{sec:EfficSubMatrix}

Let $P, N \in \mathbbm{N}$, where $P$ is prime.  DeVore, using techniques along the lines of Kashin \cite{KashinDetCS}, gives a deterministic construction of $\left( P, \left\lfloor \frac{\ln N}{\ln P} \right\rfloor \right)$-coherent matrices having $P^2$ rows and $N$ columns in \cite{devore2007deterministic}.  This construction, together with Bertrand's postulate, yields general $\left( K, \left\lfloor \frac{\ln N}{\ln K} \right\rfloor \right)$-coherent matrices with $K^2 \leq m < 4K^2$ rows and $N$ columns for any given $K, N \in \mathbbm{N}$.  
In section~\ref{sec:SampMat} we will start to construct compressed sensing matrices of near-optimal size by randomly selecting a small set of rows from one of DeVore's deterministic $\left( K, \left\lfloor \frac{\ln N}{\ln K} \right\rfloor \right)$-coherent matrices.  However, before doing so we will discuss the complexity of storing and regenerating submatrices of DeVore's $\left( K, \left\lfloor \frac{\ln N}{\ln K} \right\rfloor \right)$-coherent matrices.

As above, let $P, N \in \mathbbm{N}$ with $P$ prime.  Furthermore, let $F$ be the finite field of order $P$.  Every column of a $P^2 \times N$ $\left( P, \left\lfloor \frac{\ln N}{\ln P} \right\rfloor \right)$-coherent matrix as constructed in \cite{devore2007deterministic} has an associated polynomial over $F$ of degree at most $\lceil \log_P N \rceil - 1$.  We will assume that these polynomials are ordered so that the polynomial associated $j^{\rm th}$-column is 
$$Q_j(x) := j_0 + j_1 x + j_2 x^2 + \dots + j_{\lceil \log_P N \rceil - 1} x^{\lceil \log_P N \rceil - 1},$$ 
where $j_0, \dots, j_{\lceil \log_P N \rceil - 1} \in [P]$ are the digits of $j \in [N]$ base $P$.  That is, 
$$j = j_0 + j_1 P + j_2 P^2 + \dots + j_{\lceil \log_P N \rceil - 1} P^{\lceil \log_P N \rceil - 1}.$$
Thus, $Q_j$ with $j \in [N]$ can be obtained in $O(\log_P N)$-time by finding the representation of $j$ base $P$.\footnote{We assume O(1)-time arithmetic operations (e.g., $+$, $-$, $\cdot$, $/$) throughout this paper.}  

Let $M \in \{ 0, 1 \}^{P^2 \times N}$ be a $\left( P, \left\lfloor \frac{\ln N}{\ln P} \right\rfloor \right)$-coherent matrix as constructed in \cite{devore2007deterministic}.  The $P^2$ rows of $M$ are indexed by elements of $[P] \times [P]$, ordered lexicographically.  Given $j \in [N]$ the ones in the $j^{\rm th}$-column of $M$ appear in rows $$\left(0,Q_j(0) \right),\left(1,Q_j(1) \right), \left(2,Q_j(2) \right), \dots, \left(P-1, Q_j(P-1) \right).$$
Given $p \in [P]$ we will refer to the set of $P$ rows of $M$, 
$$\left\{ \left(p,r \right) ~\big|~ r \in [P] \right\},$$
as the \textit{$p^{\rm th}$ block of rows}.  Every column of $M$, $j \in [N]$, will have exactly one $1$ in each such block.  Furthermore, this $1$ can be located in $O(\log_P N)$-time by using Horner's rule on $Q_j$.

\section{Existence of Near Optimal $\left( K, \alpha \right)$-coherent Matrices}
\label{sec:OptExistence}

In this section we will use standard probabilistic arguments to demonstrate the existence of $\left( K, \alpha \right)$-coherent matrices having a near-optimal number of rows.  In particular, we will demonstrate that a randomly generated matrix having $m = O \left( \frac{K^2}{\alpha} \right)$ rows will be $\left( K, \alpha \right)$-coherent with high probability.  The end result is that the methods herein can be utilized as the basis for a $O(m N)$-time Monte Carlo algorithm for building near-optimal $\left( K, \alpha \right)$-coherent matrices.\footnote{Furthermore, it is worth recalling that these same methods also provide Las Vegas algorithms which run in expected $O(m N^2)$-time.}

DeVore's construction yields $\left( K, \left\lfloor \frac{\ln N}{\ln K} \right\rfloor \right)$-coherent matrices having $O(K^2)$ rows, exceeding the lower bound, $\Omega \left( \frac{K^2}{\alpha^2} \log_{K / \alpha} N \right)$, by an $\alpha = O(\log_{K} N)$-factor (assuming that $K \gg \alpha$).  In this section we demonstrate the existence of $\left( \Theta(K), \Theta(\ln N ) \right)$-coherent matrices having $O(K^2 / \alpha)$ rows.  These matrices exceed the lower bound by a $O(\log K)$ factor, and represent a general improvement over DeVore's construction with respect to row count. 

We will build $M \in \{ 0,1 \}^{m \times N}$ by letting each entry, $M_{i,j}$, be an independent and identically distributed Bernoulli random variable that is $1$ with probability $p$ and $0$ with probability $1 - p$.  Note that the number of ones in a given column of $M$ will be a binomial random variable in this case.  Similarly, the inner product between any two given columns of $M$ will also be binomial.  Hence, we may bound both of these quantities using the Chernoff and union bounds.  We have the following two lemmas.



\begin{lem}
Let $\sigma, p \in [0,1)$ and $m,N,K \in \mathbbm{N}$.  Randomly generate a matrix, $M \in \{ 0,1 \}^{m \times N}$, each of whose entries is an i.i.d. Bernoulli random variable which is $1$ with probability $p$.  If $K$ is $\Omega(\log N)$ and
\begin{equation}
mp ~=~ K + \ln \left( \frac{3N}{1 - \sigma} \right) + \sqrt{ 2 K \ln \left( \frac{3N}{1 - \sigma} \right) + \ln^2 \left( \frac{3N}{1 - \sigma} \right)}
\label{equ:MPset}
\end{equation}
then every column of $M$ will contain $\Theta(K)$ ones with probability at least
$1 - \frac{2(1 - \sigma)}{3}$. 
\label{lem:Rand_Cond1}
\end{lem}

\noindent \textit{Proof:}  Let $S_j$, $j \in [N]$, be the number of ones in column $j$ of $M$.  We have $\mathbbm{E} \left[ S_j \right] = m p$.  The Chernoff bound now implies that
$$\mathbbm{P} \left[ S_j < K \right] = \mathbbm{P}\left[ S_j < \frac{K}{m p}\cdot \mathbbm{E} \left[ S_j \right] \right] < e^{-mp \left(1-\frac{K}{m p}\right)^2 / 2}$$
as long as $K < mp$.  Thus, we can bound the probability that $S_j < K$ above by $\frac{1 - \sigma}{3N}$ for any desired $\sigma \in [0,1)$ by ensuring that
\begin{equation}
\mathbbm{P} \left[ S_j < K \right] < e^{-mp \left(1-\frac{K}{m p}\right)^2 / 2} \leq \frac{1 - \sigma}{3N}.
\label{equ:Rand_Cond1}
\end{equation}

Simplifying Equation~\ref{equ:Rand_Cond1} above, we obtain
$$mp \left(1-\frac{K}{m p}\right)^2 \geq 2 \ln \left( \frac{3N}{1 - \sigma} \right).$$
Solving for $mp$ in terms of $K$ and $N$, we learn that
$$\left( mp \right)^2 - 2 \left( K + \ln \left( \frac{3N}{1 - \sigma} \right) \right)\left( mp \right) + K^2 ~\geq~ 0.$$
This will hold whenever
$$mp ~\geq~ K + \ln \left( \frac{3N}{1 - \sigma} \right) + \sqrt{ 2 K \ln \left( \frac{3N}{1 - \sigma} \right) + \ln^2 \left( \frac{3N}{1 - \sigma} \right)} ~>~ K.$$
Applying Equation~\ref{equ:Rand_Cond1} together with the union bound over all $N$ choices of $S_j$ yields the desired lower bound.  A similar 
argument guarantees that every row will also have fewer than
$$eK + e \ln \left( \frac{3N}{1 - \sigma} \right) + e \sqrt{ 2 K \ln \left( \frac{3N}{1 - \sigma} \right) + \ln^2 \left( \frac{3N}{1 - \sigma} \right)}$$ 
ones with probability at least $1 - \frac{1 - \sigma}{3}$.~~$\Box$\\

\begin{lem}
Let $\sigma, p \in [0,1)$ and $m,N,K \in \mathbbm{N}$.  Randomly generate a matrix, $M \in \{ 0,1 \}^{m \times N}$, each of whose entries is an i.i.d. Bernoulli random variable which is $1$ with probability $p$.  If $\alpha = 2mp^2 \geq 2 \log_{4/e} \left( \frac{3N^2}{1 - \sigma} \right)$ then all pairs of columns of $M$ will have inner product at most $\alpha$ with probability at least $1 - \frac{1 - \sigma}{3}$. 
\label{lem:Rand_Cond2}
\end{lem}

\noindent \textit{Proof:}  Let $I_{i,j}$ be the inner product of the $j^{\rm th}$ column of $M$ with the $i^{\rm th}$ column of $M$ for a given $i,j \in [N]$ with $i \neq j$.  We want $I_{i,j} \leq \alpha$.  Since $I_{i,j}$ is binomial with $\mathbbm{E} \left[ I_{i,j} \right] = m p^2$ the Chernoff bound implies that
$$\mathbbm{P} \left[ I_{i,j} > \alpha \right] = \mathbbm{P}\left[ I_{i,j} > \frac{\alpha}{m p^2}\cdot \mathbbm{E} \left[ I_{i,j} \right] \right] < \left[ \frac{e^{\alpha / m p^2}}{e \left( \frac{\alpha}{m p^2} \right)^{\alpha / m p^2}} \right]^{m p^2}$$
as long as $\alpha > mp^2$.  For the sake of simplicity, suppose that $\alpha = 2mp^2 = 2 \log_{4/e} \left( \frac{3N^2}{1 - \sigma} \right)$.  Then,
$$\mathbbm{P} \left[ I_{i,j} > \alpha \right] < \left( \frac{e}{4} \right)^{\log_{4/e} \left( \frac{3N^2}{1 - \sigma} \right)} = \frac{1 - \sigma}{3N^2}.$$
In this case the union bound now guarantees that our randomly constructed matrix $M$ will also satisfy the second $\left( K, \alpha \right)$-coherent property with probability at least $1 - \frac{1 - \sigma}{3}$.~~$\Box$\\

Lemma~\ref{lem:Rand_Cond1} guarantees that a randomly constructed binary matrix will satisfy the first $\left( K, \alpha \right)$-coherent property with high probability.  Similarly, Lemma~\ref{lem:Rand_Cond2} guarantees the second $\left( K, \alpha \right)$-coherent property.  Solving for $p$ in light of Equation~\ref{equ:MPset} and Lemma~\ref{lem:Rand_Cond2} we get that we can set
\begin{equation}
p = \frac{m p^2}{ mp} = \frac{\log_{4/e} \left( \frac{3N^2}{1 - \sigma} \right)}{K + \ln \left( \frac{3N}{1 - \sigma} \right) + \sqrt{ 2 K \ln \left( \frac{3N}{1 - \sigma} \right) + \ln^2 \left( \frac{3N}{1 - \sigma} \right)}}
\label{equ:Prob}
\end{equation}
and
\begin{equation}
m = \frac{mp}{p} = \frac{\Bigg( K + \ln \left( \frac{3N}{1 - \sigma} \right) + \sqrt{ 2 K \ln \left( \frac{3N}{1 - \sigma} \right) + \ln^2 \left( \frac{3N}{1 - \sigma} \right)} \Bigg)^2}{\log_{4/e} \left( \frac{3N^2}{1 - \sigma} \right)}.
\label{equ:Rows}
\end{equation}
Note these equations both make sense whenever $K \geq \alpha \geq2 \log_{4/e} \left( \frac{3N^2}{1 - \sigma} \right)$.  We have the following.

\begin{thm}
Fix $\sigma \in [0,1)$.  Let $m, K, \alpha \in [N]$ be such that $K \geq \alpha \geq2 \log_{4/e} \left( \frac{3N^2}{1 - \sigma} \right)$, and let $m = \Theta \left( K^2 / \alpha \right)$ as per Equation~\ref{equ:Rows}.  Randomly generate a matrix, $M \in \{ 0,1 \}^{m \times N}$, each of whose entries is an i.i.d. Bernoulli random variable which is $1$ with the probability, $p$, given in Equation~\ref{equ:Prob}.  Then, $M$ will be both $\left( K, \alpha \right)$-coherent and have $\Theta \left( K \right)$ ones per column with probability at least $\sigma$.
\label{thm:OptM_Exist}
\end{thm}

Although the matrices developed above have fewer rows than DeVore's, we hasten to point out that they are generally less structured.  This ultimately means that they will be difficult to store in compact form, and, therefore, of limited use when space complexity is a dominant concern.

\section{Sampling Rows from a $\left( K, \alpha \right)$-coherent Matrix}
\label{sec:SampMat}

Given an $m \times N$ matrix $M$ and a subset $s \subset [m]$, we define $M_{s}$ to be the $|s| \times N$ sub matrix of $M$ consisting of the rows of $M$ contained in $s$.  If $s$ is explicitly specified to be a multiset as opposed to a set, we will (implicitly) repeat rows from $M$ contained in $s$ as necessary.  Let $l_n$ denote the number of nonzero entries in the $n^{\rm th}$ column of $M_{s}$.  We define $M_{s}(l,n)$, $1 \leq l \leq l_n$, to be the $l \times N$ sub matrix of  $M_{s}$ consisting of the first $l$ rows in $M_{s}$ with nonzero entries in the $n^{th}$ column.

\subsection{Identification Matrix}

The following corollary to Lemma~\ref{applem:Mest} will be used to construct matrices for the identification of the largest magnitude entries in ${\bf x}$.  Note that the corollary is essentially a coupon collection result (i.e., we want to collect, for each element in $S^{\rm opt}_{2k / \epsilon}$, a ``good row'' satisfying Equation~\ref{eqn:PropRand_Loc}).

\begin{cor}
Suppose $M$ is an $m \times N$ $\left(K, \alpha \right)$-coherent matrix.  Let $\epsilon^{-1} \in \mathbbm{N}^+ $, $k \in \left[ \epsilon K / \alpha \right]$, $c \in [14,\infty) \cap \mathbbm{N}$, $\sigma \in [2/3,1)$, and ${\bf x} \in \mathbbm{R}^{N}$.  Select a subset of the rows of $M$, $s' \subset [m]$, by independently choosing 
\begin{equation}
\gamma ~\geq~ \frac{7}{6} \cdot \frac{m}{K} \ln \left( \frac{2k / \epsilon }{1-\sigma} \right)
\label{eqn:Rand_rows_Loc}
\end{equation}
values from $[m]$ uniformly at random with replacement.  If $K > c \cdot (k \alpha / \epsilon)$ then with probability at least $\sigma$ every $n \in S^{\rm opt}_{2k / \epsilon} \subset [N]$ will have an associated row of $M_{s'}$, $i_n \in [ \gamma ]$, for which
\begin{equation}
\left| \left(M_{s'} \cdot {\bf x} \right)_{i_n} - x_n \right| \leq \frac{\epsilon \cdot \left\| {\bf x} - {\bf x}^{\rm opt}_{(k/\epsilon)} \right\|_1}{k}.
\label{eqn:PropRand_Loc}
\end{equation}
\label{cor:RandM_Loc}
\end{cor}

\noindent \textit{Proof:}  Fix $n \in S^{\rm opt}_{2k / \epsilon}$.  Lemma~\ref{applem:Mest} implies that each randomly selected row of $M$, $j \in [m]$, will satisfy Equation~\ref{eqn:PropRand_Loc} with probability at least $\frac{6}{7} \cdot \frac{K}{m}$.  Hence, the probability that none of the $\gamma$ selected rows will satisfy Equation~\ref{eqn:PropRand_Loc} is at most
$$\left( 1 - \frac{6}{7} \cdot \frac{K}{m} \right)^\gamma.$$

Let $x = \frac{7}{6} \cdot \frac{m}{K}.$  If $\gamma$ satisfies Equation~\ref{eqn:Rand_rows_Loc} we have that
$$\gamma \left( 1 + \sum^{\infty}_{h = 2} \frac{1}{h x^{h-1}} \right) \geq x \cdot \ln \left( \frac{2k / \epsilon }{1-\sigma} \right).$$
This in turn implies that 
$$\gamma \left( \sum^{\infty}_{h = 1} \frac{1}{h x^{h}} \right) ~=~ - \gamma \cdot \ln \left(1 - \frac{1}{x} \right) \geq \ln \left( \frac{2k / \epsilon}{1-\sigma} \right).$$
Thus, 
$$\left( 1 - \frac{6}{7} \cdot \frac{K}{m} \right)^\gamma ~\leq~ \frac{1-\sigma}{2k / \epsilon}$$
whenever $\gamma$ satisfies Equation~\ref{eqn:Rand_rows_Loc}.  Taking the union bound over all $2k / \epsilon$ elements of $S^{\rm opt}_{2k / \epsilon}$ finishes the proof.~~$\Box$\\

It is straightforward to show that a random sub matrix, $M_{s'}$, will have $O(\log N)$ ones in every column with high probability when it is constructed as per Corollary~\ref{cor:RandM_Loc} from a $\left(K, \alpha \right)$-coherent matrix having $\Theta \left( K \right)$ ones per column.\footnote{This result follows via techniques analogous to those utilized in Section~\ref{sec:OptExistence} in order to establish Theorem~\ref{thm:OptM_Exist} (i.e., via the Chernoff and union bounds).}  It is also important to note that an analogous variant of  Corollary~\ref{cor:RandM_Loc} can be proven for DeVore's $\left( \Theta(K), \Theta(\log_K N) \right)$-coherent matrices by randomly selecting $O \left( \ln \left( \frac{2k / \epsilon }{1-\sigma} \right) \right)$  \textit{blocks of $\Theta(K)$ rows} (see Section~\ref{sec:EfficSubMatrix}).  Randomly selecting rows from a DeVore matrix in blocks both $(i)$ guarantees that every column of the resulting sub matrix will have $O \left( \ln \left( \frac{2k / \epsilon }{1-\sigma} \right) \right)$ ones, and $(ii)$ requires only $O \left( \ln \left( \frac{2k / \epsilon }{1-\sigma} \right) \ln K \right)$ random bits.  The following theorem is proven via standard bit testing techniques (see, e.g., \cite{cormode2006combinatorial,SIAM_Fourier_Matrix}).

\begin{thm}
Suppose $M$ is an $m \times N$ $\left(K, \alpha \right)$-coherent matrix.  Let $\epsilon \in (0,1]$, $\sigma \in [2/3,1)$, $k \in \left[K \cdot \frac{\epsilon}{14 \alpha} \right]$, and ${\bf x} \in \mathbbm{R}^{N}$.  Construct $M_{s'}$ as per Corollary~\ref{cor:RandM_Loc}.  Then, with probability at least $\sigma$, $\left(M_{s'} \circledast \mathcal{B}_N \right) {\bf x}$ will allow Phase 1 (i.e., lines 4 through 14) of Algorithm~\ref{appalg:reconstruct} in the appendix to recover all $n \in [N]$ for which
\begin{equation}
\left| x_n \right| \geq 4\frac{\epsilon \cdot \left\| {\bf x} - {\bf x}^{\rm opt}_{(k/\epsilon)} \right\|_1}{k}.
\label{equ:BigXn}
\end{equation}
The required Phase 1 runtime is $O \left(\frac{m}{K} \ln \left( \frac{2k / \epsilon }{1-\sigma} \right) \ln N \right)$.
\label{thm:Identify}
\end{thm}

The only new observation required for the proof of Theorem~\ref{thm:Identify} beyond those used to prove the analogous results in \cite{cormode2006combinatorial,SIAM_Fourier_Matrix,Iwen2012} involves noting that any $n$ satisfying Equation~\ref{equ:BigXn} also belongs to $S^{\rm opt}_{2k / \epsilon}$.

To finish, we note that applying (a variant of) Corollary~\ref{cor:RandM_Loc} to a $\left( \Theta(K), \Theta(\log_K N) \right)$-coherent matrix from Section~\ref{sec:EfficSubMatrix} produces a random matrix, $M_{s'}$, having 
$$O \left(K \ln \left( \frac{2k / \epsilon }{1-\sigma} \right) \right) = O \left(\frac{k}{\epsilon} \cdot \log_{k/\epsilon} N \cdot \ln \left( \frac{2k / \epsilon }{1-\sigma} \right) \right)$$ 
rows.  For $\sigma$ fixed, this reduces to $O \left( (k / \epsilon) \log N \right)$ rows.  Furthermore, $M_{s'}$ will have $O \left( \log( k / \epsilon) \right)$ ones in all columns.  Applying Corollary~\ref{cor:RandM_Loc} to a $\left( \Theta(K), \Theta(\log N) \right)$-coherent matrix from Section~\ref{sec:OptExistence} produces a random matrix, $M_{s'}$, having 
$$O \left(\frac{m}{K} \ln \left( \frac{2k / \epsilon }{1-\sigma} \right) \right) = O \left(\frac{K}{\log N} \cdot \ln \left( \frac{2k / \epsilon }{1-\sigma} \right) \right)  = O \left(\frac{k}{\epsilon} \cdot \ln \left( \frac{2k / \epsilon }{1-\sigma} \right) \right)$$ 
rows.  For $\sigma$ fixed, this reduces to $O \left( (k / \epsilon) \log( k / \epsilon) \right)$ rows.  Furthermore, $M_{s'}$ will have $O \left( \log N \right)$ ones in all columns with high probability.  

\subsection{Estimation Matrix}

The following corollary constructs measurements capable of estimating every entry of ${\bf x}$ that is identified as large in magnitude during Phase 1 of Algorithm~\ref{appalg:reconstruct}. Furthermore, the estimation procedure is simple, requiring only median operations (see Phase 2 of Algorithm~\ref{appalg:reconstruct}).  

\begin{cor}
Suppose $M$ is an $m \times N$ $\left(K, \alpha \right)$-coherent matrix.  Let $\epsilon^{-1} \in \mathbbm{N}^+ $, $k \in \left[ \epsilon K / \alpha \right]$, $c \in [14,\infty) \cap \mathbbm{N}$, $\sigma \in [2/3,1)$, $S \subseteq [N]$, and ${\bf x} \in \mathbbm{R}^{N}$.  Select a multiset of the rows of $M$, $\tilde{s} \subset [m]$, by independently choosing 
\begin{equation}
\beta \geq 28.56 \cdot \frac{m}{K} \ln \left( \frac{2 |S|}{1-\sigma} \right)
\label{eqn:Rand_rows}
\end{equation}
values from $[m]$ uniformly at random with replacement.  If $K > c \cdot (k \alpha / \epsilon)$ then $M_{\tilde{s}}$ will have both of the following properties with probability at least $\sigma$:  
\begin{enumerate}
\item There will be at least $\tilde{l} = 21 \cdot \ln \left( \frac{2 |S|}{1-\sigma} \right)$ nonzero values in every column of $M_{\tilde{s}}$ indexed by $S$.  Hence, $M_{\tilde{s}}(\tilde{l},n)$ will be well defined for all $n \in S$.
\item For all $n \in S$ more than $l_n / 2$ of the entries in $M_{\tilde{s}}(l_n,n) \cdot {\bf x}$ ~(i.e., more than half of the values $j \in [l_n]$, counted with multiplicity) will have
$$\left| \left(M_{\tilde{s}}(l_n,n) \cdot {\bf x} \right)_j - x_n \right| \leq \frac{\epsilon \cdot \left\| {\bf x} - {\bf x}^{\rm opt}_{(k/\epsilon)} \right\|_1}{k}.$$
\end{enumerate}
\label{cor:RandM}
\end{cor}

\noindent \textit{Proof:}  Fix $n \in S$.  We select our multiset, $\tilde{s} \subset [m]$, of the rows of $M$ by independently choosing $\beta$ elements of $[m]$ uniformly at random with replacement.  Denote the $j^{\rm th}$ element chosen for $\tilde{s}$ by $\tilde{s}_j$.  Finally, let $P^n_j$ be the random variable indicating whether $M_{\tilde{s}_j,n} > 0$, and let $Q^n_j$ be the random variable indicating whether $\tilde{s}_{j}$ satisfies
\begin{equation}
\left| \left(M \cdot {\bf x}\right)_{\tilde{s}_j} - x_n \right| \leq \frac{\epsilon \cdot \left\| {\bf x} - {\bf x}^{\rm opt}_{(k/\epsilon)} \right\|_1}{k}
\label{eqn:PropRand}
\end{equation}
conditioned on $P^n_j$.  Thus, $P^n_j = 1$ if $M_{\tilde{s}_j,n} > 0$, and $0$ otherwise.  Similarly, 
$$Q^n_j = \left\{ \begin{array}{ll} 1 & \textrm{if } \tilde{s}_{j} \textrm{ satisfies Equation~\ref{eqn:PropRand} and } P^n_j = 1\\ 0 & {\rm otherwise} \end{array} \right..$$
Lemma~\ref{applem:Mest} implies that $\mathbbm{P}\left[ Q^n_j = 1 ~\big|~ P^n_j = 1 \right] > \frac{6}{7}$.  Furthermore, 
$$\mu = \mathbbm{E} \left[ \sum^{\beta}_{j = 1} Q^n_j ~\big|~ P^n_1, \dots, P^n_{\beta} \right] \geq \frac{6}{7} \left( \sum^{\beta}_{j = 1} P^n_j \right).$$

Let $l_n = \sum^{\beta}_{j = 1} P^n_j$.  The Chernoff bound (see, e.g., \cite{raghavan1995randomized}) guarantees that 
$$\mathbbm{P}\left[\sum^{\beta}_{j = 1} Q^n_j < \frac{4 \cdot l_n}{7} ~\bigg|~ l_n \right] \leq e^{- \frac{\mu}{18}} \leq e^{-\frac{l_n}{21}}.$$  
Thus, if $l_n > 21$ we can see that $\sum^{\beta}_{j = 1} Q^n_j$ will be less than $\frac{l_n + 1}{2}$ with probability less than $e^{-\frac{l_n}{21}}$.  Hence, if $l_n \geq 21 \ln \left( \frac{2|S|}{1-\sigma} \right)$ then Property~2 will fail to be satisfied for $n$ with probability less than $\frac{1-\sigma}{2|S|}$.  Focusing now on $l_n$, we note that $\mathbbm{P}\left[ P^n_j = 1 \right] \geq \frac{K}{m}$ so that $\tilde{\mu} = \mathbbm{E} \left[ l_n \right] \geq \frac{K}{m} \beta$.

Let $\tilde{l} = 21 \ln \left( \frac{2|S|}{1-\sigma} \right)$. Applying the Chernoff bound one additional time reveals that $\mathbbm{P}\left[ l_n ~<~ \tilde{l} \right] < \mathbbm{e}^{- \tilde{\mu} \cdot \left( 1 - \frac{\tilde{l}}{\tilde{\mu}}  \right)^2 / 2 }.$  Hence, if we wish to bound $\mathbbm{P}\left[ l_n ~<~ \tilde{l} \right]$ from above by $\frac{1-\sigma}{2|S|}$ it suffices to have $\tilde{\mu}^2 - \frac{44}{21} \tilde{\mu} \tilde{l} + \tilde{l}^2 \geq 0$.  Setting $\beta \geq 1.36 \cdot \frac{m}{K} \tilde{l} = 28.56 \cdot \frac{m}{K} \ln \left( \frac{2|S|}{1-\sigma} \right)$ achieves this goal.  The end result is that $M_{\tilde{s}}$ will fail to satisfy both Properties~1 and~2 for any $n \in S$ with probability less than $\frac{1-\sigma}{|S|}$.  Applying the union bound over all $n \in S$ finishes the proof.~~$\Box$\\

Note that corollary~\ref{cor:RandM} considers selecting a multiset of rows from a $\left(K, \alpha \right)$-coherent matrix.  Hence, some rows may be selected more than once.  If this occurs, rows should be considered to be selected multiple times for counting purposes only.  That is, all computations involving a row which is selected several times should still be carried out only once.  However, the results of these computations should be considered with greater weight during subsequent reconstruction efforts (e.g., multiplely selected rows should be considered as generating multiple duplicate entries in $M_{\tilde{s}} \cdot {\bf x}$).  


As above, it is straightforward to show that a random sub matrix, $M_{\tilde{s}}$, will have $O(\log N)$ ones in every column with high probability when it is constructed as per Corollary~\ref{cor:RandM} from a $\left(K, \alpha \right)$-coherent matrix having $\Theta \left( K \right)$ ones per column.  In addition, an analogous variant of  Corollary~\ref{cor:RandM} can be proven for DeVore's $\left( \Theta(K), \Theta(\log_K N) \right)$-coherent matrices by randomly selecting $O \left(\ln \left( \frac{2 |S|}{1-\sigma} \right) \right)$ blocks of $\Theta(K)$ rows.  Randomly selecting rows from a DeVore matrix in blocks this way both guarantees that all columns of the resulting sub matrix will have $O \left(\ln \left( \frac{2 |S|}{1-\sigma} \right) \right)$ ones, and also requires only $O \left(\ln \left( \frac{2 |S|}{1-\sigma} \right) \ln K \right)$ random bits.  Note that we must be able to quickly construct arbitrary columns of $M_{\tilde{s}}$ in order to execute Phase 2 of Algorithm~\ref{appalg:reconstruct} in the low memory setting (i.e., when we can not explicitly store either the entire matrix $M$, or the randomly selected sub matrix $M_{\tilde{s}}$ in memory).  In this setting DeVore's $\left( \Theta(K), \Theta(\log_K N) \right)$-coherent matrices allow us to reconstruct any column of a random sub matrix containing $O \left(\ln \left( \frac{2 |S|}{1-\sigma} \right) \right)$ blocks of rows, $M_{\tilde{s}}$, in just $O\left(\ln \left( \frac{2 |S|}{1-\sigma} \right) \cdot \log_K N \right)$-time (see Section~\ref{sec:EfficSubMatrix} for details).

Corollary~\ref{cor:RandM} will generally be applied with $S \subset [N]$ set to the subset discovered by Phase 1 of Algorithm~\ref{appalg:reconstruct}.\footnote{In fact, we select the rows for $M_{\tilde{s}}$ independently of the subset, $S$, found during Phase 1 of Algorithm~\ref{appalg:reconstruct}, \textit{before} the subset has been identified.  Note that we only require an upper bound on the size of $S$ before selecting rows from $M$ for our estimation matrix.  Such an upper bound is supplied in advance by Corollary~\ref{cor:RandM_Loc}.}  Hence, we will generally have $|S|$ equal to the number of rows in a matrix $M_{s'}$ constructed via Corollary~\ref{cor:RandM_Loc}.  In more extreme settings, where we want to be able to estimate \textit{all} entries of ${\bf x}$ with high probability, we will set $S = [N]$.  Corollary~\ref{cor:RandM} implies the following theorem.

\begin{thm}
Suppose $M$ is an $m \times N$ $\left(K, \alpha \right)$-coherent matrix.  Let $\epsilon \in (0,1]$, $\sigma \in [2/3,1)$, $k \in \left[K \cdot \frac{\epsilon}{14 \alpha} \right]$, and ${\bf x} \in \mathbbm{R}^{N}$.  Construct $M_{\tilde{s}}$ as per Corollary~\ref{cor:RandM}.  Then, with probability at least $\sigma$, $M_{\tilde{s}} {\bf x}$ will allow Phase 2 (i.e., lines 15 through 19) of Algorithm~\ref{appalg:reconstruct} in the appendix to estimate all $x_n$ with $n \in S$ with a $z_n$ satisfying
\begin{equation}
\left| z_n - x_n \right| \leq \frac{\epsilon \cdot \left\| {\bf x} - {\bf x}^{\rm opt}_{(k/\epsilon)} \right\|_1}{k}.
\label{equ:ErrorBound}
\end{equation}
The required Phase 2 runtime (and memory complexity) is $O\left(|S|  \ln \left( \frac{2 |S|}{1-\sigma} \right)  \cdot \log_K N \right)$ when $M$ is a $\left( \Theta(K), \Theta(\log_K N) \right)$-coherent DeVore matrix.  Phase 2 requires $O\left(|S| \cdot \log N \right)$-time if $M$ is a $\left( \Theta(K), \Theta(\log N) \right)$-coherent matrix from Section~\ref{sec:OptExistence}.
\label{thm:Estimate}
\end{thm}

\noindent \textit{Proof:}  Equation~\ref{equ:ErrorBound} follows from the second property of $M_{\tilde{s}}$ guaranteed by Corollary~\ref{cor:RandM}.  Lines 15 through 17 can be accomplished in $O\left(|S| \cdot \ln \left( \frac{2 |S|}{1-\sigma} \right) \cdot \log_K N \right)$-time using a median-of-medians algorithm when $M$ is a $\left( \Theta(K), \Theta(\log_K N) \right)$-coherent DeVore matrix.  When $M$ is a $\left( \Theta(K), \Theta(\log N) \right)$-coherent matrix from Section~\ref{sec:OptExistence}, lines 15 through 17 can be accomplished in $O\left(|S| \cdot \log N \right)$-time.\footnote{However, using the matrixes from Section~\ref{sec:OptExistence} requires $O(N \log N)$-memory since their columns contain ones in random locations that must be remembered.}  Lines 18 and 19 can always be accomplished in $O(|S| \log |S|)$-time. ~~$\Box$\\

We conclude this section by noting that applying (a variant of) Corollary~\ref{cor:RandM} to a $\left( \Theta(K), \Theta(\log_K N) \right)$-coherent matrix from Section~\ref{sec:EfficSubMatrix} produces a random matrix, $M_{\tilde{s}}$, having 
$$O \left(K \ln \left( \frac{2 |S|}{1-\sigma} \right) \right) = O \left(\frac{k}{\epsilon} \cdot \log_{k/\epsilon} N \cdot \ln \left( \frac{2 |S|}{1-\sigma} \right)  \right)$$ 
rows.  Applying Corollary~\ref{cor:RandM} to a $\left( \Theta(K), \Theta(\log N) \right)$-coherent matrix from Section~\ref{sec:OptExistence} produces a random matrix, $M_{\tilde{s}}$, having 
$$O \left(\frac{m}{K} \ln \left( \frac{2 |S|}{1-\sigma} \right) \right) = O \left(\frac{K}{\log N} \cdot \ln \left( \frac{2 |S|}{1-\sigma} \right) \right)  = O \left(\frac{k}{\epsilon} \cdot \ln \left( \frac{2 |S|}{1-\sigma} \right) \right)$$ 
rows.  For $\sigma$ fixed, this reduces to $O \left( (k / \epsilon) \log( |S| ) \right)$ rows.  Furthermore, $M_{\tilde{s}}$ will have $O \left( \log N \right)$ ones in all columns with high probability.  

\section{Main Results}
\label{sec:MainRes}

We may now prove the three new results mentioned in Section~\ref{sec:IntroResandRel}.  We have the following theorem.

\begin{thm}
Let $\epsilon \in (0,1]$, $\sigma \in [2/3,1)$, ${\bf x} \in \mathbbm{R}^{N}$, and $k \in [N]$.\footnote{For the sake of simplicity, we assume $k = \Omega(\log N)$ when stating the measurement and runtime bounds below.}  With probability at least $\sigma$ Algorithm~\ref{appalg:reconstruct} will output a vector ${\bf z} \in \mathbbm{R}^N$ satisfying
\begin{equation}
\left\| {\bf x} - {\bf z} \right\|_2 ~\leq~ \left\| {\bf x} - {\bf x}^{\rm~ opt}_k \right\|_2 + \frac{22 \epsilon \left\| {\bf x} - {\bf x}^{\rm opt}_{(k/\epsilon)} \right\|_1}{\sqrt{k}}
\label{equ:ApproxError}
\end{equation}
when executed using any of the following identification and estimation matrices:
\begin{enumerate}
\item  A $\left( \Theta(k \log N / \epsilon), \Theta(\log N) \right)$-coherent matrix from Section~\ref{sec:OptExistence} used for estimation via Corollary~\ref{cor:RandM} with $S = [N]$.  Only Phase 2 of Algorithm~\ref{appalg:reconstruct} need be applied (i.e., no identification will be performed).  The resulting number of measurements is $O \left(\frac{k}{\epsilon} \cdot \ln \left( \frac{N}{1-\sigma} \right) \right)$.  The required runtime is $O(N \log N)$.
\item A $\left( \Theta(\frac{k}{\epsilon} \log_{k/ \epsilon} N), \Theta(\log_{k / \epsilon} N) \right)$-coherent matrix from Section~\ref{sec:EfficSubMatrix} used for both identification (via Corollary~\ref{cor:RandM_Loc} variant) and estimation (via Corollary~\ref{cor:RandM} variant with $|S| = O(\frac{k}{\epsilon} \ln \left( N / (1-\sigma) \right) )$).  The resulting number of measurements is $O \left(\frac{k}{\epsilon} \cdot \ln \left( \frac{N}{1-\sigma} \right)  \ln N \right)$.  The required runtime is $O \left(\frac{k}{\epsilon} \cdot \ln^2 \left( \frac{N}{1-\sigma} \right) \right)$.
\item A $\left( \Theta(k \log N / \epsilon), \Theta(\log N) \right)$-coherent matrix from Section~\ref{sec:OptExistence} used for identification (via Corollary~\ref{cor:RandM_Loc}), and a $\left( \Theta(\frac{k}{\epsilon} \log_{k/ \epsilon} N), \Theta(\log_{k / \epsilon} N) \right)$-coherent matrix from Section~\ref{sec:EfficSubMatrix} used for estimation (via Corollary~\ref{cor:RandM} variant with $|S| = O(\frac{k}{\epsilon} \ln \left( k / \epsilon (1-\sigma) \right) )$).  The resulting number of measurements is $O \left(\frac{k}{\epsilon} \cdot \ln \left( \frac{k / \epsilon }{1-\sigma} \right) \ln N \right)$.  The required runtime is $O \left(\frac{k}{\epsilon} \cdot \ln \left( \frac{k / \epsilon }{1-\sigma} \right) \ln \left( \frac{N}{1-\sigma} \right) \right)$.
\end{enumerate}
\label{appthm:NewRecoverAlg}
\end{thm}

\noindent \textit{Proof:}  The runtime and measurement bounds follow from Theorem~\ref{thm:Identify}, Theorem~\ref{thm:Estimate}, and the subsequent Section~\ref{sec:SampMat} discussions.  The error guarantee for ${\bf z}$ follows from Theorem~\ref{thm:Identify}, Theorem~\ref{thm:Estimate}, and the proof of Theorem 7 in \cite{Iwen2012}.
~~$\Box$\\

It is interesting to consider the possibility of improving the runtime bounds obtained in Theorem~\ref{appthm:NewRecoverAlg} by using iterative recovery techniques akin to those employed in \cite{gilbert2010approximate}.  This appears to be difficult.  In particular, such iterative recovery methods generally require the contributions of partial solutions to be subtracted from the input measurements of the original vector, ${\bf x}$, after each of $O(\log k)$ rounds.  Assuming that one must subtract some partial solution containing at least $\Omega(k)$ nonzero entries from a large (constant fraction) of the initial measurements of ${\bf x}$ at some point during reconstruction, it becomes clear that updating our measurements will not be $O(k \log N)$-time unless our measurement matrix contains $O(\log N)$ nonzero entries per column.  Unfortunately, fast nonadaptive identification of previously undiscovered heavy elements of ${\bf x}$ (e.g., via bit-testing methods) requires the use of matrices having $\Omega(\log N)$ nonzero entries in many columns during each new round of iterative approximation.  Hence, it appears as if only $O(1)$ rounds of identification may be performed using the techniques considered herein before the required measurement matrices have too many ones per column in order to allow $O(k \log N)$-time recovery.  The author considers this as (a weak) justification for utilizing only one round of identification in Algorithm~\ref{appalg:reconstruct}.

\section{Conclusion}
\label{sec:Conc}

In this paper we present a compressed sensing recovery algorithm with an ``$\ell_2,\ell_1$'' error guarantee that runs in only $O\left( (k \log k ) \log N \right)$-time.  This runtime is within a $O(\log k)$ factor of the known lower $\Omega(k \log N)$ runtime bound.  Demonstrating (or refuting) the existence of a $O(k \log N)$-time (i.e., linear-time in its required input size) compressed sensing recovery algorithm with similar error guarantees remains an open problem.  


\bibliographystyle{abbrv}
\bibliography{SODA13}

\appendix

\section{The Recovery Algorithm}

See Algorithm~\ref{appalg:reconstruct} below.

\begin{algorithm}[tb]
\begin{algorithmic}[1]
\caption{$\proc{Approximate} ~{\bf x}$} 
\label{appalg:reconstruct}
\STATE \textbf{Input:  $M_{\tilde{s}}$ and $M_{\tilde{s}} {\bf x}$ for estimation, and $\left(M_{s'} \circledast \mathcal{B}_N\right) {\bf x}$ for identification} 
\STATE \textbf{Output: ${\bf z}$, an approximation to } ${\bf x}^{\rm~ opt}_k$
\STATE Initialize multiset $S \leftarrow \emptyset, ~{\bf z} \leftarrow {\bf 0} \in \mathbbm{R}^N, ~{\bf b} \leftarrow {\bf 0} \in \mathbbm{R}^{\lceil \log_2 N \rceil}$
\begin{center}
{\sc Phase 1:  Identify All Heavy $n \in [0,N) \cap \mathbbm{N}$}
\end{center}
\FOR {$j$ from $1$ to $|s'|$}
	\FOR {$i$ from $1$ to $\lceil \log_2 N \rceil$}
		\IF {$\left| \left( M_{s'} \circledast \left(\mathcal{B}_N \right)_{i+1} {\bf x} \right)_j \right| ~>~ \left| \left( M_{s'} {\bf x} ~-~ M_{s'} \circledast \left(\mathcal{B}_N \right)_{i+1} {\bf x} \right)_j \right|$}
			\STATE $b_i \leftarrow 1$
		\ELSE
			\STATE $b_i \leftarrow 0$
		\ENDIF
	\ENDFOR
	\STATE $n \leftarrow \sum^{\lceil \log_2 N \rceil-1}_{i=0} b_{i+1} 2^i$
	\STATE $S \leftarrow S \cup \{ n \}$
\ENDFOR \\
\begin{center}
{\sc Phase 2:  Estimate ${\bf x}_S \approx {\bf x}^{\rm~ opt}_k$ Using Equation~\ref{eqn:CombProp}}
\end{center}
\FOR {\textbf{each} $n$ value belonging to $S$}
	\STATE $z_n \leftarrow \textrm{median~of~multiset} \left\{ \left(M_{\tilde{s}}(l_n,n) \cdot {\bf x}_{} \right)_h~\big|~1 \leq h \leq l_n \right\}$
\ENDFOR
\STATE Sort nonzero ${\bf z}$ entries by magnitude so that $|z_{n_1}| \geq |z_{n_2}| \geq |z_{n_3}| \geq \dots$
\STATE $S \leftarrow \{ n_1, n_2, \dots, n_{2k} \}$
\STATE Output: ${\bf z}_{S}$
\end{algorithmic}
\end{algorithm}





\end{document}